# Electronic properties of metamorphic GaSbBi films on GaAs

*Joonas Hilska,[1(a)] Janne Puustinen,[1] Eero Koivusalo,[1] Mircea Guina[1]*

[1] Optoelectronics Research Centre, Physics Unit, Tampere University, FI-33720 Tampere, Finland

(a) Corresponding author's e-mail: joonas.hilska@tuni.fi

We report on the electronic, structural, and optical properties of epitaxial GaSbBi films with varying Bi-concentration (up to 7%Bi) grown on semi-insulating GaAs(100) substrates. The 1 µm thick GaSbBi epilayers exhibit fully relaxed narrow X-ray diffraction peaks and smooth surface morphology comparable to that of high-quality GaSb epilayers on GaAs. Low temperature photoluminescence spectra exhibit band gap shrinkage consistent with Bi alloying. Electrical Hall measurements indicate reduction of hole concentration and no change in the hole mobilities with increasing Bi content for the nominally undoped GaSbBi alloy. The residual hole concentration reduces from $10^{18}$ cm$^{-3}$ level for a reference GaSb sample, to low $10^{17}$ cm$^{-3}$ level with increasing Bi content. Hole mobility values of around 300 cm$^2$/Vs are observed independent of the Bi content. These dependencies are attributed to the Bi surfactant effect and Bi-induced defect formation.

**Introduction**

Advanced III-V heterostructures utilizing GaSbBi alloys have important application potential in mid-infrared optoelectronics, thermoelectric generators and thermophotovoltaic devices [1–4]. Consequently, increasing research efforts have been dedicated to develop epitaxial growth of high quality GaSbBi alloys with high Bi content while focusing on investigating their structural and optical properties [5–11]. However, comparatively little attention has been paid to investigating the electrical properties of these novel alloys, which are of key importance for the performance in end-use devices. In fact, one can find only cursory studies examining GaSbBi alloys with very low Bi content (<1%Bi) fabricated by liquid phase epitaxy [12, 13] and a study on GaSbBi Schottky diodes with up to 5%Bi grown by molecular beam epitaxy (MBE) [14] with the focus solely on excess hole concentration. This lack of studies on electrical properties of GaSbBi alloys originates from the unavailability of commercial semi-insulating (SI) GaSb substrates as well as from the requirement of growing thick homogeneous GaSbBi epilayers, which would enable simple van der Pauw geometry Hall measurements. Therefore, it is highly interesting to explore the growth of these alloys on foreign substrates like GaAs, which enable the use of non-conductive substrates for Hall experiments. Moreover, growing on GaAs instead of GaSb substrates provides a more economical and scalable technology platform for application deployment (we note that GaAs substrates are the cheapest and have the largest diameter amongst all III-V alternatives). To this end, we demonstrate the fabrication of 1 µm thick GaSbBi epilayers by MBE on SI-GaAs(100). The focus is on revealing the interplay between their structural, optical and electrical properties and the Bi content (up to ~7%Bi, which is deemed high enough for significant band-gap engineering capability). In contrast to the early studies on electrical properties of GaSbBi [14], we demonstrate that GaSbBi alloys can in fact be grown with significant Bi content while maintaining hole concentration in the low $10^{17}$ cm$^{-3}$ level. We furthermore show that high hole mobilities around ~300 cm$^2$/Vs can be achieved independent of the Bi content. We examine these observations in context of Bi's ability to act as a surfactant and the tendency to form specific point defects typical for III-V-Bi alloys.

## Experimental methodology

The investigated GaSbBi samples were grown on SI-GaAs(100) substrates by solid-source MBE using standard effusion cells for Al, Ga, and Bi, and valved cracker sources for Sb and As. The sample layer structure is shown in Fig. 1(a), wherein the composition of the top 1 µm thick layer is varied between samples. The underlying layer structure consist of an undoped GaAs buffer, a 2 nm AlSb interfacial misfit (IMF) layer and an intermediate 10 nm GaSb layer. The IMF layer facilitates lattice relaxation at the GaAs/AlSb interface through formation of a periodic 90° dislocation array [15, 16]. Five GaSbBi samples labelled B1 through B5 were grown with increasing Bi content in the range of ~1-7%Bi. We also fabricated two GaSb reference samples labelled R1 and R2. The sample R1 was grown under typical GaSb growth conditions with V/III atomic flux ratio of ~1.7 and at a temperature of 520 °C. The sample R2 was grown under the same conditions as all the Bi-containing samples, namely V/III flux ratio of ~1.0 and at a low temperature (LT) of 300 °C. The growth temperatures reported are estimated by pyrometry and true atomic flux ratios are calibrated by Sb cap and GaAs sample growths. The Bi content for samples B1-B5 was controlled by tuning the Bi flux while the Ga flux was kept at a constant value for all samples corresponding to a growth rate of ~0.8 µm/h. The structural and optical properties are characterized with X-ray diffraction (XRD), atomic force microscopy (AFM) and low temperature photoluminescence (LT-PL). Electrical properties were measured via Hall effect at room temperature with the van der Pauw method.

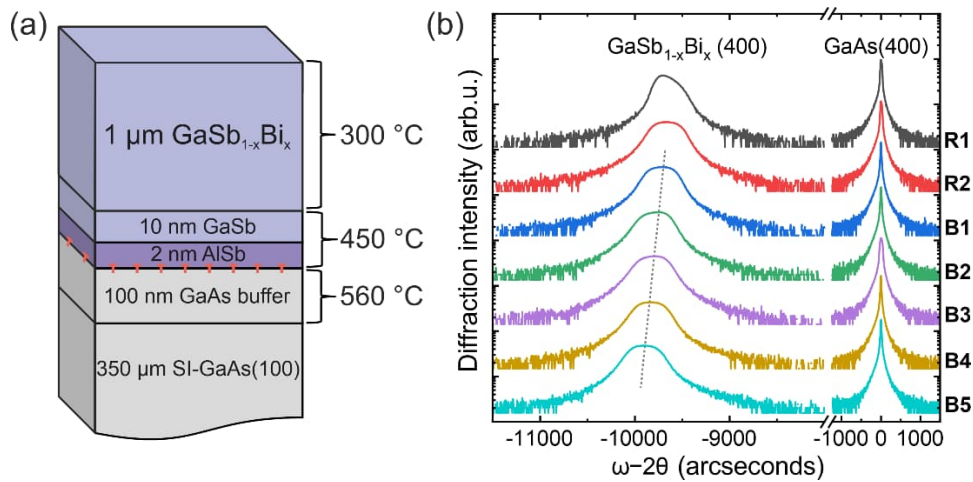

**FIG. 1.** (a) Layer structure of the investigated samples with the growth temperatures indicated on the right-hand side. For sample R1, the GaSb layer was grown at 520 °C. (b) XRD ω-2θ measurements centered to the GaAs(400) diffraction peak. The dashed line over the $GaSb_{1-x}Bi_x$ diffraction peaks indicates the increase of the lattice constant with increasing Bi content.

Fig. 1(b) comprises XRD ω-2θ measurements for all samples and reveals two strong diffraction peaks corresponding to the GaAs substrate and the GaSb(Bi) epilayers. The peak separation around ~9700 arcsec for samples R1 and R2 corresponds to ~99% relaxed GaSb layers. The essentially complete relaxation was confirmed by reciprocal space mapping for all samples with representative measurements given in Fig. S1 of the supplementary information. For the Bi-containing epilayers, the peak separation grows with increasing Bi content for samples B1-B5 as indicated by the dashed line in Fig. 1(b). We estimate the Bi content from this increasing mismatch by assuming 100% relaxation and estimating the GaSbBi lattice constant from Vegard's law using lattice constants of 6.0954 Å and 6.27 Å for GaSb [17] and GaBi [6]. The GaSb(Bi) diffraction peak full width at half-maximum (FWHM) values are consistently ~300 arcsec for the low temperature grown reference R2 and B1-B5 samples. The high temperature reference R1 exhibits the lowest FWHM at around 200 arcsec matching the lowest values in literature for similarly grown GaSb-on-GaAs [15]. The slight asymmetry in the GaSb peak of sample R1 likely originates from dilute As-incorporation from the residual As-pressure after GaAs buffer growth, together with enhanced Sb-to-As anion exchange process at high growth temperature [18].

Surface morphology of selected samples is compiled in Fig. 2. For sample R1 in Fig. 2(a), the morphology follows that found in literature for high temperature grown GaSb-on-GaAs [15, 19], exhibiting three dimensional growth with clear monolayer terraces and sub-nanometer root mean square (RMS) roughness of 0.69 nm. For sample R2 in Fig. 2(b), the surface morphology is altered to morphology commonly observed in LT-grown GaSb [20] and GaSbBi alloys [6, 8, 21] grown on native GaSb substrates. This morphology arises from enhanced Ehrlich–Schwoebel energy barriers at low temperature, which inhibit downward movement of atoms across step-edges [22]. The RMS roughness for R2 is 1.43 nm, still well within common roughness values for GaSb-on-GaAs growth [15]. For the Bi-containing samples B1 (1.1%Bi) and B4 (4.7%Bi), the morphology is similar to R2, however the mound-period increases and the mounds elongate anisotropically. The RMS roughness is similar at ~1.3 nm for both samples B1 and B4. For two of the higher Bi content samples B3 (3.4%Bi) and B5 (6.7%Bi), small Bi droplets are observed on the surface (cf. supplementary information Fig. S2). This is due to narrower growth window in the V/III-ratio for droplet-free growth as the Bi content is increased [8]. Deviation by a percentage point or two from the ideal V/III ratio can result into significant accumulation of excess Bi atoms on the growth front, given the large thickness of the epilayer.

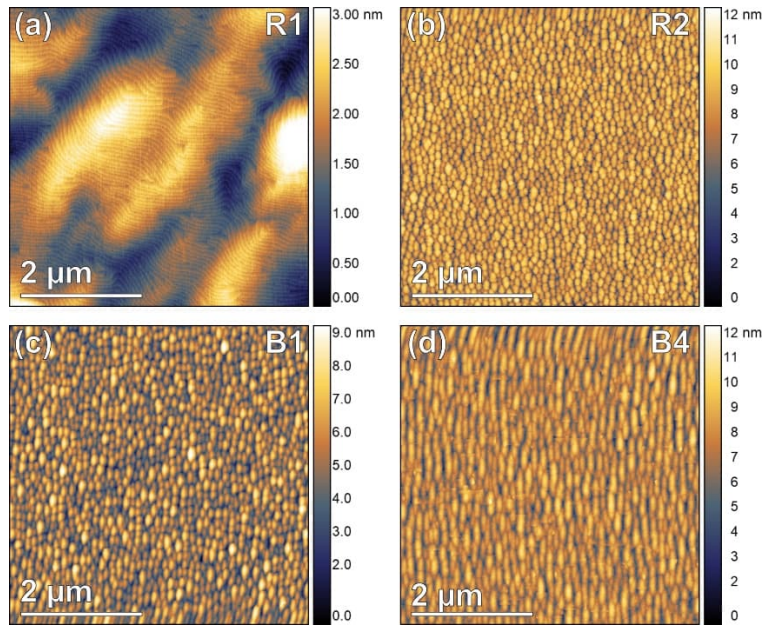

**FIG. 2.** (a)-(d) 5×5 µm$^2$ AFM scans for GaSb samples R1 (a) and R2 (b) and two GaSb$_{1-x}$Bi$_x$ samples, B1 (c) and B4 (d).

Fig. 3 comprises LT-PL spectra corresponding to samples B1-B5. The samples exhibit clear PL emission with wavelength shifting to higher values as the Bi content is increased. Despite the use of metamorphic LT-growth, the samples exhibit relatively strong emission. This strong emission is an indirect indication of a low threading dislocation density (~10$^6$-10$^8$ cm$^{-2}$ [15]) in combination with the thin higher band gap GaSb layer blocking carriers from the dislocation array at the GaAs/AlSb interface. Emission wavelengths range from 1.8 μm to 2.4 μm, which corresponds to a band gap shift of -33 meV/%Bi, consistent with reported values for GaSbBi grown on GaSb [9, 21, 23]. Sample B4 exhibits a secondary peak on the short wavelength side. Such multicomponent emission is commonly observed for GaSbBi [9] as well as other III-V-bismides [24–27], which is typically ascribed to shallow localized defect states and/or alloy inhomogeneity.

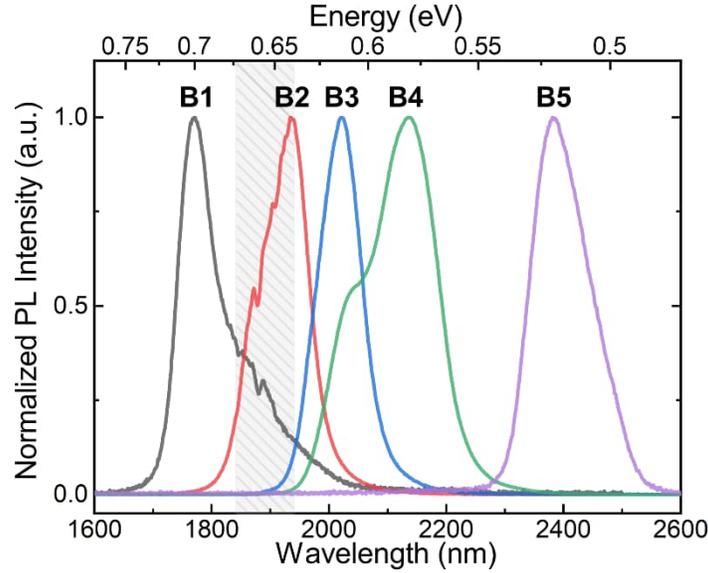

**FIG. 3.** Normalized LT-PL spectra from samples B1-B5 measured at 12 K with 532 nm excitation. The shaded area around 1900 nm is influenced by atmospheric water absorption along the optical path of the setup. The intensities are normalized due to measurements being performed with different optical filters.

The electrical characterization is summarized in Fig. 4 with values also tabulated in Table I together with data from AFM and XRD measurements. Hole mobility for the high temperature GaSb sample R1 is 620 cm$^2$/Vs with a hole concentration of 4.7×10$^{16}$ cm$^{-3}$, which correspond to values found for high-quality MBE grown GaSb-on-GaAs [28–32]. In contrast, the low temperature GaSb reference R2 mobility is around half at 275 cm$^2$/Vs and the hole concentration is around 20-times larger at 9.5×10$^{17}$ cm$^{-3}$. Significant decrease in hole concentration is observed for the Bi-containing samples when compared to the reference sample R2 grown under the same conditions. Compared to the reference sample R2, the hole concentration for B1 (1.1%Bi) decreases to around half, to 5×10$^{17}$ cm$^{-3}$. With further increase of the Bi content the hole concentration reduces to around one-tenth from the reference value (R2) to a concentration around (1-2)×10$^{17}$ cm$^{-3}$ (samples B2-B5), and remains relatively constant for different Bi-fractions.

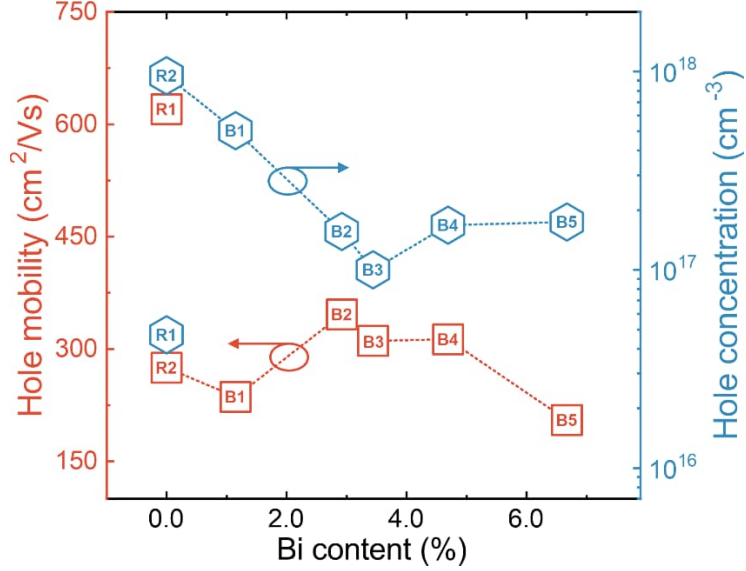

**FIG. 4.** Hole mobility (left; red squares) and hole concentration (right; blue hexagons) as determined by Hall measurements at room temperature. Datapoints for samples grown at low temperature are connected by dashed lines.

**Table I.** Summary of all the samples in this study.

| Sample | Bi content (%) | $T_{growth}$ (°C) | (400) FWHM (arcsec) | Mobility (cm$^2$/Vs) | Hole concentration (×10$^{16}$ cm$^{-3}$) | Roughness[a] (nm) |
|---|---|---|---|---|---|---|
| **R1** | 0 | 520 | 210 | 620 | 4.7 | 0.69 |
| **R2** | 0 | 300 | 300 | 275 | 95.4 | 1.43 |
| **B1** | 1.1 | 300 | 295 | 236 | 50.3 | 1.30 |
| **B2** | 2.9 | 300 | 302 | 346 | 15.6 | 1.20 |
| **B3** | 3.4 | 300 | 307 | 311 | 10.0 | 11.03* |
| **B4** | 4.7 | 300 | 308 | 313 | 16.8 | 1.30 |
| **B5** | 6.7 | 300 | 308 | 205 | 17.4 | 4.99* |

[a]Root mean square roughness value for an AFM scan area of 5×5 µm$^2$.
*Samples with Bi droplets on the surface, cf. supplementary information Fig. S2.

## Discussion

It is well understood that the origin of native p-type doping in GaSb is governed by Ga$_{Sb}$ antisite and V$_{Ga}$ vacancy defect formation [33], with the latter indicated to be more significant defect in terms of concentration for MBE-grown GaSb [34]. We therefore examine the results of the electrical measurements in the context of these defects, the defects potentially introduced by Bi and the changed growth kinetics. For example, comparing R1 to R2, we can understand the 20-times larger hole concentration in the low temperature reference being due to reduced Ga surface diffusion rates, which can be clearly seen in the changed surface morphology in Fig. 2 (a-b). Namely, as the Ga diffusion is reduced, the likelihood of Ga atoms locating correct group III sublattice sites for incorporation is also reduced. Thus, the likelihood of Ga-atoms forming Ga$_{Sb}$ antisites and $V_{Ga}$ vacancies is increased, and hole concentration increased from R1 to R2. The high defect concentration is also reflected in the significant mobility decrease through the carriers interacting with the high density of ionized defects.

Once Bi is introduced, we observe a clear hole concentration decrease from sample R1 to saturated concentration around $10^{17}$ cm$^{-3}$ for samples B2-B5. This decrease can be either due to (i) reduction of the native $Ga_{Sb}$ antisite and $V_{Ga}$ defect concentration in the alloy by Bi, or (ii) compensation of the native acceptor states by Bi-related defect donor states. The former could be understood by the Bi surfactant effect increasing the mean Ga diffusion rates over the surface and reducing hole concentration by the logic outlined above. We can qualitatively examine the plausibility of this explanation by analysis of the surface defect density in sample R2. The hole concentration or equivalently the Ga-related point defect concentration of R2 is ~$10^{18}$ cm$^{-3}$ translating to a surface defect density of $(10^{18}/10^{22})\times5.4\times10^{14}$ cm$^{-2}$≈$5\times10^{10}$ cm$^{-2}$. This is equivalent to square area of ~50×50 nm$^2$ per defect, which corresponds half of the surface mound periodicity of R2 in Fig. 2(b) (see also Fig. S3 in the supplementary information). By comparison to Fig. 2 (c-d), the mound periodicity is increased for the Bi-containing samples, meaning Ga surface diffusion rates are likely increased and thus hole concentration is reduced.

The Bi surfactant effect has been widely reported for various III-Vs [35–37] as well as for GaSb [38]. In fact, the Bi surfactant mediated reduction in defect density has been identified for (In)GaNAs [39] and InGaAs [40]. The saturation of this effect could perhaps be interpreted as saturated surfactant coverage limiting its effect on the Ga diffusion length, and thus minimum hole concentration. We also point out that despite samples B3 and B5 having higher surface roughness due to Bi droplets (cf. Table I and the Fig. S2 of the supplementary information), they do not deviate from the hole concentration trend of the sample set. This is due to the Bi droplet density being three orders of magnitude lower than the surface defect density, meaning they do not play a significant role in the point defect formation process.

To discuss the effects of acceptor compensation via Bi-induced donor defects, we examine commonly suggested point defects in III-V-Bi alloys. Undoped GaAsBi is natively p-type and exhibits Bi content dependent hole concentrations increasing up to $10^{17}$ cm$^{-3}$ for ~10%Bi [41]. The acceptor state origin has been tentatively suggested to be due to high-order $Bi_n$ (n>3) clustering in the group V sublattice, since lower order $Bi_n$ (n≤3) clustering is expected to be much more concentrated and strongly dependent on Bi content for randomly distributed Bi atoms in the lattice [42]. This interpretation would be consistent for hole concentration trend observed by Segercrantz et al. for GaSbBi [14], where increasing hole concentrations up to $10^{19}$ cm$^{-3}$ were observed for ~5%Bi. However completely opposite type of Bi-induced electronic states have been observed for Be or C doped p-GaAsBi [43], wherein clear acceptor compensation was observed for increasing Bi content whose origin was ascribed to heteroantisite $Bi_{Ga}$ formation. Similar Bi induced donor states are consistent for observations in InGaAsBi [44]. In fact, our earlier investigation of Raman scattering from GaSbBi evidenced pure $Bi_2$ cluster formation [45], i.e. cluster involving Bi-Bi bonding where one Bi atom on a group V site is neighbored with a $Bi_{Ga}$ hetero-antisite, that makes us to think that such defects are playing a role for our observations.

Pure $Bi_2$ cluster defect formation has been theoretically shown to be energetically favorable around $V_{Ga}$ vacancies in GaAsBi [46]. The $V_{Ga}$ concentration is linked to the hole concentration [33, 34], which for the sample R2 is very high (~$10^{18}$ cm$^{-3}$). Thus, upon Bi incorporation in such alloy with high $V_{Ga}$ concentration implies high likelihood of $Bi_2$ cluster formation [46]. Furthermore, the thermally generated Bi flux consists of roughly equal parts of $Bi_1$ monomers and $Bi_2$ dimers

[47], which could offer simple incorporation pathway of these clusters through non-dissociative $Bi_2$ incorporation. However, within the scope of our study we cannot rule out the possibility of isolated $Bi_{Ga}$ heteroantisite formation or any defect complex which would have the same compensating effects observed here. The observation of the $Bi_{Ga}$ defect has been a long-standing quest for GaAsBi with mixed findings on its existence [48–50]. Very recently, comprehensive tunneling electron microscopy investigation of GaAsBi/AlGaAs quantum wells excluded any other reason for observed anomalous contrast than $Bi_{Ga}$ [51]. In general, the Bi related defect landscape for different III-V-Bi alloys, and even the same alloys grown under different conditions, is complex.

Finally, in terms of the hole mobility we do not observe any clear degradation with increasing Bi content. This is in stark contrast of observations in GaAsBi, wherein the hole mobility clearly degrades with increasing Bi [41, 43]. A part of this disparity could be ascribed to the non-highly-mismatched alloy nature of GaSbBi in comparison to GaAsBi, as pointed out by Luna et al. [52, 53]. Namely, as there is less significant difference in Sb versus Bi in size and/or electronegativity, the Bi induced perturbations in the valence band structure degrading mobility are less significant for GaSbBi in comparison to GaAsBi. However, the relatively constant hole mobility observed here could also be explained by interplay of both defect reduction by the surfactant effect and compensation by Bi-induced defects resulting in a zero-sum influence on the mobility.

## Conclusions

We studied the properties of MBE grown thick GaSbBi epilayers with varying Bi contents (up to ~7%Bi) on SI-GaAs(100) substrates. XRD and AFM measurements indicated crystalline quality similar to GaSb-on-GaAs growth, smooth surfaces ($R_q$ ~1 nm), and homogeneous Bi incorporation. Low temperature PL exhibited clear band edge related emission with band gap shrinkage consistent with increasing Bi alloy fraction of -33 meV/%Bi. Most notably, Hall measurements showed that the hole concentration in the GaSbBi alloys can be kept comparably low despite high Bi content in the alloy, around $2\times10^{17}$ $cm^{-3}$ level for ~7%Bi. This is two orders of magnitude less than indicated in an earlier report [14], demonstrating feasibility of GaSbBi in device heterostructures requiring low background doping. We also report high GaSbBi hole mobilities of around ~300 $cm^2$/Vs, which is independent of Bi content.

## Acknowledgements

Financial support from the Finnish Research Council projects QuantSi (decision no. 323989), CryoLight (decision no. 357351), and Flagship Programme PREIN is acknowledged.

## References

1. Jung, D., Bank, S., Lee, M.L., Wasserman, D.: Next-generation mid-infrared sources. J. Opt. 19, 123001 (2017). https://doi.org/10.1088/2040-8986/aa939b

2. Sweeney, S.J., Eales, T.D., Marko, I.P.: The physics of mid-infrared semiconductor materials and heterostructures. In: Mid-infrared Optoelectronics. pp. 3–56. Elsevier (2020)

3. Delorme, O., Cerutti, L., Luna, E., Narcy, G., Trampert, A., Tournié, E., Rodriguez, J.-B.: GaSbBi/GaSb quantum well laser diodes. Applied Physics Letters. 110, 222106 (2017). https://doi.org/10.1063/1.4984799

4. AlFaify, S., Ul Haq, B., Ahmed, R., Butt, F.K., Alsardia, M.M.: Investigation of $GaBi_{1-x}Sb_x$ based highly mismatched alloys: Potential thermoelectric materials for renewable energy devices and applications. Journal of Alloys and Compounds. 739, 380–387 (2018). https://doi.org/10.1016/j.jallcom.2017.12.306

5. Rajpalke, M.K., Linhart, W.M., Birkett, M., Yu, K.M., Scanlon, D.O., Buckeridge, J., Jones, T.S., Ashwin, M.J., Veal, T.D.: Growth and properties of GaSbBi alloys. Applied Physics Letters. 103, 142106 (2013). https://doi.org/10.1063/1.4824077

6. Rajpalke, M.K., Linhart, W.M., Birkett, M., Yu, K.M., Alaria, J., Kopaczek, J., Kudrawiec, R., Jones, T.S., Ashwin, M.J., Veal, T.D.: High Bi content GaSbBi alloys. Journal of Applied Physics. 116, 043511 (2014). https://doi.org/10.1063/1.4891217

7. Delorme, O., Cerutti, L., Tournié, E., Rodriguez, J.-B.: Molecular beam epitaxy and characterization of high Bi content GaSbBi alloys. Journal of Crystal Growth. 477, 144–148 (2017). https://doi.org/10.1016/j.jcrysgro.2017.03.048

8. Hilska, J., Koivusalo, E., Puustinen, J., Suomalainen, S., Guina, M.: Epitaxial phases of high Bi content GaSbBi alloys. Journal of Crystal Growth. 516, 67–71 (2019). https://doi.org/10.1016/j.jcrysgro.2019.03.028

9. Kopaczek, J., Kudrawiec, R., Linhart, W., Rajpalke, M., Jones, T., Ashwin, M., Veal, T.: Low- and high-energy photoluminescence from GaSb$_{1-x}$Bi$_x$ with 0<x≤0.042. Appl. Phys. Express. 7, 111202 (2014). https://doi.org/10.7567/APEX.7.111202

10. Smołka, T., Rygała, M., Hilska, J., Puustinen, J., Koivusalo, E., Guina, M., Motyka, M.: Influence of the Bismuth Content on the Optical Properties and Photoluminescence Decay Time in GaSbBi Films. ACS Omega. 8, 36355–36360 (2023). https://doi.org/10.1021/acsomega.3c05046

11. McElearney, J., Grossklaus, K., Menasuta, T.P., Vandervelde, T.: Determination of the Complex Refractive Index of GaSb$_{1-x}$Bi$_x$ by Variable-Angle Spectroscopic Ellipsometry. Physica Status Solidi (a). 221, 2400017 (2024). https://doi.org/10.1002/pssa.202400017

12. Sharma, A.S., Das, S., Gazi, S.A., Dhar, S.: Influence of Pb vs Ga solvents during liquid phase epitaxy on the optical and electrical properties of GaSbBi layers. Journal of Applied Physics. 126, 155702 (2019). https://doi.org/10.1063/1.5120754

13. Sharma, A.S., Sreerag, S.J., Kini, R.N.: Temperature-dependent ultrafast hot carrier dynamics in the dilute bismide alloy GaSb$_{1-x}$Bi$_x$ (x≲ 0.4%). Journal of Applied Physics. 135, 035701 (2024). https://doi.org/10.1063/5.0179135

14. Segercrantz, N., Slotte, J., Makkonen, I., Tuomisto, F., Sandall, I.C., Ashwin, M.J., Veal, T.D.: Hole density and acceptor-type defects in MBE-grown GaSb$_{1-x}$Bi$_x$. J. Phys. D: Appl. Phys. 50, 295102 (2017). https://doi.org/10.1088/1361-6463/aa779a

15. Jasik, A., Ratajczak, J., Sankowska, I., Wawro, A., Smoczyński, D., Czuba, K.: LT-AlSb Interlayer as a Filter of Threading Dislocations in GaSb Grown on (001) GaAs Substrate Using MBE. Crystals. 9, 628 (2019). https://doi.org/10.3390/cryst9120628

16. Huang, S.H., Balakrishnan, G., Khoshakhlagh, A., Jallipalli, A., Dawson, L.R., Huffaker, D.L.: Strain relief by periodic misfit arrays for low defect density GaSb on GaAs. Applied Physics Letters. 88, 131911 (2006). https://doi.org/10.1063/1.2172742

17. Vurgaftman, I., Meyer, J.R., Ram-Mohan, L.R.: Band parameters for III–V compound semiconductors and their alloys. Journal of Applied Physics. 89, 5815–5875 (2001). https://doi.org/10.1063/1.1368156

18. Xie, Q., Van Nostrand, J.E., Brown, J.L., Stutz, C.E.: Arsenic for antimony exchange on GaSb, its impacts on surface morphology, and interface structure. Journal of Applied Physics. 86, 329–337 (1999). https://doi.org/10.1063/1.370733

19. Jasik, A., Sankowska, I., Wawro, A., Ratajczak, J., Smoczyński, D., Czuba, K.: GaSb layers with low defect density deposited on (001) GaAs substrate in two-dimensional growth mode using molecular beam epitaxy. Current Applied Physics. 19, 542–547 (2019). https://doi.org/10.1016/j.cap.2019.02.012

20. Nosho, B.Z., Bennett, B.R., Aifer, E.H., Goldenberg, M.: Surface morphology of homoepitaxial GaSb films grown on flat and vicinal substrates. Journal of Crystal Growth. 236, 155–164 (2002). https://doi.org/10.1016/S0022-0248(01)02392-2

21. Rajpalke, M.K., Linhart, W.M., Yu, K.M., Jones, T.S., Ashwin, M.J., Veal, T.D.: Bi flux-dependent MBE growth of GaSbBi alloys. Journal of Crystal Growth. 425, 241–244 (2015). https://doi.org/10.1016/j.jcrysgro.2015.02.093

22. Apostolopoulos, G., Boukos, N., Herfort, J., Travlos, A., Ploog, K.H.: Surface morphology of low temperature grown GaAs on singular and vicinal substrates. Materials Science and Engineering: B. 88, 205–208 (2002). https://doi.org/10.1016/S0921-5107(01)00905-9

23. Kopaczek, J., Kudrawiec, R., Linhart, W.M., Rajpalke, M.K., Yu, K.M., Jones, T.S., Ashwin, M.J., Misiewicz, J., Veal, T.D.: Temperature dependence of the band gap of GaSb$_{1-x}$Bi$_x$ alloys with 0<x≤0.042 determined by photoreflectance. Applied Physics Letters. 103, 261907 (2013). https://doi.org/10.1063/1.4858967

24. Francoeur, S., Tixier, S., Young, E., Tiedje, T., Mascarenhas, A.: Bi isoelectronic impurities in GaAs. Phys. Rev. B. 77, 085209 (2008). https://doi.org/10.1103/PhysRevB.77.085209

25. Mazzucato, S., Lehec, H., Carrère, H., Makhloufi, H., Arnoult, A., Fontaine, C., Amand, T., Marie, X.: Low-temperature photoluminescence study of exciton recombination in bulk GaAsBi. Nanoscale Res Lett. 9, 19 (2014). https://doi.org/10.1186/1556-276X-9-19

26. Bahrami-Yekta, V., Tiedje, T., Masnadi-Shirazi, M.: MBE growth optimization for GaAs$_{1-x}$Bi$_x$ and dependence of photoluminescence on growth temperature. Semicond. Sci. Technol. 30, 094007 (2015). https://doi.org/10.1088/0268-1242/30/9/094007

27. Wu, X., Chen, X., Pan, W., Wang, P., Zhang, L., Li, Y., Wang, H., Wang, K., Shao, J., Wang, S.: Anomalous photoluminescence in InP$_{1-x}$Bi$_x$. Sci Rep. 6, 27867 (2016). https://doi.org/10.1038/srep27867

28. Jasik, A., Sankowska, I., Wawro, A., Ratajczak, J., Jakieła, R., Pierścińska, D., Smoczyński, D., Czuba, K., Regiński, K.: Comprehensive investigation of the interfacial misfit array formation in GaSb/GaAs material system. Appl. Phys. A. 124, 512 (2018). https://doi.org/10.1007/s00339-018-1931-8

29. Ivanov, S.V., Altukhov, P.D., Argunova, T.S., Bakun, A.A., Budza, A.A., Chaldyshev, V.V., Kovalenko, Y.A., Kop'ev, P.S., Kutt, R.N., Meltser, B.Y., Ruvimov, S.S., Shaposhnikov, S.V., Sorokin, L.M., Ustinov, V.M.: Molecular beam epitaxy growth and characterization of thin (<2 μm) GaSb layers on GaAs(100) substrates. Semicond. Sci. Technol. 8, 347–356 (1993). https://doi.org/10.1088/0268-1242/8/3/008

30. Xie, Q., Van Nostrand, J.E., Jones, R.L., Sizelove, J., Look, D.C.: Electrical and optical properties of undoped GaSb grown by molecular beam epitaxy using cracked Sb$_1$ and Sb$_2$.

Journal of Crystal Growth. 207, 255–265 (1999). https://doi.org/10.1016/S0022-0248(99)00379-6

31. Bosacchi, A., Franchi, S., Allegri, P., Avanzini, V., Baraldi, A., Ghezzi, C., Magnanini, R., Parisini, A., Tarricone, L.: Electrical and photoluminescence properties of undoped GaSb prepared by molecular beam epitaxy and atomic layer molecular beam epitaxy. Journal of Crystal Growth. 150, 844–848 (1995). https://doi.org/10.1016/0022-0248(95)80058-K

32. Yano, M., Suzuki, Y., Ishii, T., Matsushima, Y., Kimata, M.: Molecular Beam Epitaxy of GaSb and GaSbAs. Jpn. J. Appl. Phys. 17, 2091 (1978). https://doi.org/10.1143/JJAP.17.2091

33. Kujala, J., Segercrantz, N., Tuomisto, F., Slotte, J.: Native point defects in GaSb. Journal of Applied Physics. 116, 143508 (2014). https://doi.org/10.1063/1.4898082

34. Segercrantz, N., Slotte, J., Makkonen, I., Kujala, J., Tuomisto, F., Song, Y., Wang, S.: Point defect balance in epitaxial GaSb. Applied Physics Letters. 105, 082113 (2014). https://doi.org/10.1063/1.4894473

35. Occena, J., Jen, T., Lu, H., Carter, B.A., Jimson, T.S., Norman, A.G., Goldman, R.S.: Surfactant-induced chemical ordering of GaAsN:Bi. Applied Physics Letters. 113, 211602 (2018). https://doi.org/10.1063/1.5045606

36. M Hassanen, A., Herranz, J., Geelhaar, L., B Lewis, R.: Bismuth surfactant-enhanced III-As epitaxy on GaAs(111)A. Semicond. Sci. Technol. 38, 095009 (2023). https://doi.org/10.1088/1361-6641/ace990

37. Feng, G., Oe, K., Yoshimoto, M.: Temperature dependence of Bi behavior in MBE growth of InGaAs/InP. Journal of Crystal Growth. 301–302, 121–124 (2007). https://doi.org/10.1016/j.jcrysgro.2006.11.242

38. Menasuta, T.P., Grossklaus, K.A., McElearney, J.H., Vandervelde, T.E.: Bismuth surfactant enhancement of surface morphology and film quality of MBE-grown GaSb(100) thin films over a wide range of growth temperatures. Journal of Vacuum Science & Technology A. 42, 032703 (2024). https://doi.org/10.1116/6.0003458

39. Tixier, S., Adamcyk, M., Young, E.C., Schmid, J.H., Tiedje, T.: Surfactant enhanced growth of GaNAs and InGaNAs using bismuth. Journal of Crystal Growth. 251, 449–454 (2003). https://doi.org/10.1016/S0022-0248(02)02217-0

40. Pillai, M.R., Kim, S.-S., Ho, S.T., Barnett, S.A.: Growth of $In_xGa_{1-x}As$/GaAs heterostructures using Bi as a surfactant. Journal of Vacuum Science & Technology B: Microelectronics and Nanometer Structures Processing, Measurement, and Phenomena. 18, 1232–1236 (2000). https://doi.org/10.1116/1.591367

41. Pettinari, G., Polimeni, A., Capizzi, M., Engelkamp, H., Christianen, P.C.M., Maan, J.C., Patanè, A., Tiedje, T.: Effects of Bi incorporation on the electronic properties of GaAs: Carrier masses, hole mobility, and Bi-induced acceptor states. Physica Status Solidi (b). 250, 779–786 (2013). https://doi.org/10.1002/pssb.201200463

42. Kreitman, M.M., Barnett, D.L.: Probability Tables for Clusters of Foreign Atoms in Simple Lattices Assuming Next-Nearest-Neighbor Interactions. The Journal of Chemical Physics. 43, 364–371 (1965). https://doi.org/10.1063/1.1696753

43. Kini, R.N., Ptak, A.J., Fluegel, B., France, R., Reedy, R.C., Mascarenhas, A.: Effect of Bi alloying on the hole transport in the dilute bismide alloy GaAs$_{1-x}$Bi$_x$. Phys. Rev. B. 83, 075307 (2011). https://doi.org/10.1103/PhysRevB.83.075307

44. Petropoulos, J.P., Zhong, Y., Zide, J.M.O.: Optical and electrical characterization of InGaBiAs for use as a mid-infrared optoelectronic material. Applied Physics Letters. 99, 031110 (2011). https://doi.org/10.1063/1.3614476

45. Souto, S., Hilska, J., Galvão Gobato, Y., Souza, D., Andrade, M.B., Koivusalo, E., Puustinen, J., Guina, M.: Raman spectroscopy of GaSb$_{1-x}$Bi$_x$ alloys with high Bi content. Applied Physics Letters. 116, 202103 (2020). https://doi.org/10.1063/5.0008100

46. Punkkinen, M.P.J., Laukkanen, P., Kuzmin, M., Levämäki, H., Lång, J., Tuominen, M., Yasir, M., Dahl, J., Lu, S., Delczeg-Czirjak, E.K., Vitos, L., Kokko, K.: Does Bi form clusters in GaAs$_{1-x}$Bi$_x$ alloys? Semicond. Sci. Technol. 29, 115007 (2014). https://doi.org/10.1088/0268-1242/29/11/115007

47. Fischer, A.K.: Vapor Pressure of Bismuth. The Journal of Chemical Physics. 45, 375–377 (1966). https://doi.org/10.1063/1.1727337

48. Kunzer, M., Jost, W., Kaufmann, U., Hobgood, H.M., Thomas, R.N.: Identification of the Bi$_{Ga}$ heteroantisite defect in GaAs:Bi. Physical Review B. 48, 4437 (1993). https://doi.org/10.1103/physrevb.48.4437

49. Dagnelund, D., Puustinen, J., Guina, M., Chen, W.M., Buyanova, I.A.: Identification of an isolated arsenic antisite defect in GaAsBi. Appl. Phys. Lett. 104, 052110 (2014). https://doi.org/10.1063/1.4864644

50. Ciatto, G., Alippi, P., Bonapasta, A.A., Tiedje, T.: How much room for Bi$_{Ga}$ heteroantisites in GaAs$_{1-x}$Bi$_x$? Appl. Phys. Lett. 99, 141912 (2011). https://doi.org/10.1063/1.3647635

51. Luna, E., Puustinen, J., Hilska, J., Guina, M.: Detection of Bi$_{Ga}$ hetero-antisites at Ga(As,Bi)/(Al,Ga)As interfaces. Journal of Applied Physics. 135, 125303 (2024). https://doi.org/10.1063/5.0195965

52. Luna, E., Delorme, O., Cerutti, L., Tournié, E., Rodriguez, J.-B., Trampert, A.: Transmission electron microscopy of Ga(Sb, Bi)/GaSb quantum wells with varying Bi content and quantum well thickness. Semicond. Sci. Technol. 33, 094006 (2018). https://doi.org/10.1088/1361-6641/aad5c4

53. Luna, E., Delorme, O., Cerutti, L., Tournié, E., Rodriguez, J.-B., Trampert, A.: Microstructure and interface analysis of emerging Ga(Sb,Bi) epilayers and Ga(Sb,Bi)/GaSb quantum wells for optoelectronic applications. Applied Physics Letters. 112, 151905 (2018). https://doi.org/10.1063/1.5024199

# Supplementary Information for

# Electronic properties of metamorphic GaSbBi films on GaAs

*Joonas Hilska,*[1(a)] *Janne Puustinen,*[1] *Eero Koivusalo,*[1] *Mircea Guina*[1]

[1] Optoelectronics Research Centre, Physics Unit, Tampere University, FI-33720 Tampere, Finland

(a) Corresponding author's e-mail: joonas.hilska@tuni.fi

Figure S1. compiles XRD reciprocal space maps for sample B2 with 2.9%Bi taken from the (422) and (400) diffraction planes. In the asymmetric (422) measurement in Fig. S1 (a), the GaSbBi epilayer diffraction peak lies on the diagonal grey line drawn from the GaAs substrate peak to the reciprocal axis origin, indicating essentially 100% relaxation of the epilayer lattice. In Fig. S1 (b), the GaSbBi(400) diffraction peak is elongated along the in-plane lattice coordinate, indicating some mosaicity of the crystal.

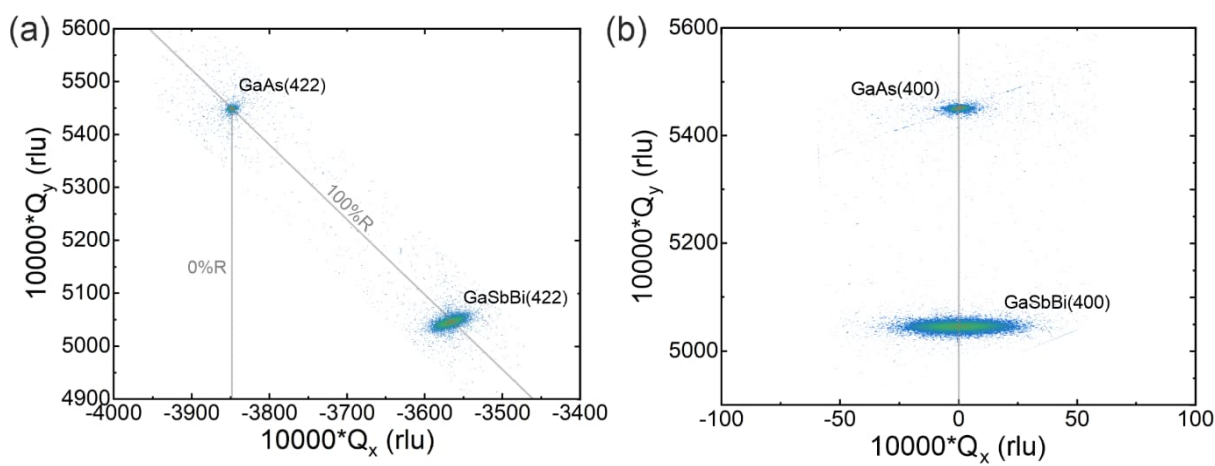

**FIG. S1.** Reciprocal space maps from the (a) (422) and (b) (400) planes for sample B2. In both figures the diffraction peaks originating from the GaAs substrate and GaSbBi epilayer are labelled. The degree of relaxation is essentially 100%.

Figure S2 comprises surface characterization for the two samples B3 and B5 with higher surface roughness (cf. **Table I**) due to Bi-droplet formation. In Fig. S2(a), the scanning electron microscope (SEM) images of sample B3 indicate isolated droplets on the surface of varying size around 100-200 nm range, shown in the secondary electron contrast image in the bottom left. These droplets deviate from spherical geometry showing some degree of faceting and in the backscattered electron detector image show up as bright dots in comparison to the background GaSbBi surface. The bright contrast is consistent with the higher atomic Z-number of the Bi atoms backscattering a larger proportion of the electron beam and the faceting is typical for Bi-droplets. In Fig. S2(b), an AFM image is shown for sample B5, which indicates again faceted deposits with background surface demonstrating undulating morphology consistent with other GaSbBi samples shown in Fig. 2 in the main text. The Bi droplets form during growth due to non-unity Bi-incorporation and low Bi desorption rates from to too high V/III flux ratio and too low surface temperature, which result in net excess Bi-accumulation on the surface forming droplets.

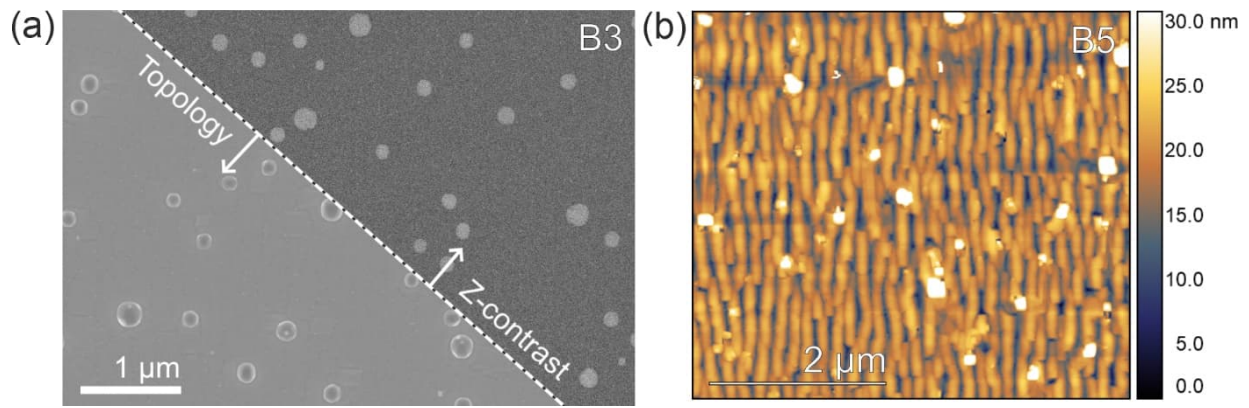

**FIG. S2.** a) Composite SEM images with backscattered electron contrast (top right) and secondary electron contrast (bottom left) showing larger Bi droplets (~100-200 nm diameter) on the surface. b) AFM 5×5 µm² scan from sample B5 showing smaller Bi droplets. Corresponding RMS roughness is 4.99 nm given in **Table I** in the main text.

Figure S3. shows a cross-sectional height profile taken from the AFM image from Fig. 2(b) from the main text. In Fig. S3, the dashed grey lines correspond to the estimated surface defect density ~50x50 nm² corresponding to the measured hole concentration for sample R2, which can be visually compared to be around half of the mound periodicity.

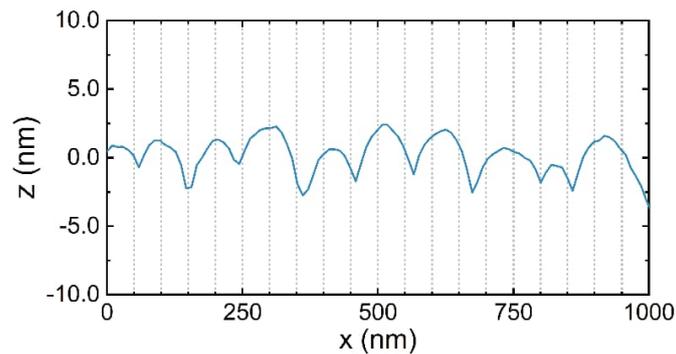

**FIG. S3.** Cross-sectional height profile from sample R2 taken from the AFM data in Fig. 2(b) in the main text. The spacing between grey dotted lines corresponds to 50 nm.